# The Ancient Astronomy of Easter Island: Mars and Sweet Potatoes

Sergei Rjabchikov[1]

[1]The Sergei Rjabchikov Foundation - Research Centre for Studies of Ancient Civilisations and Cultures, Krasnodar, Russia, e-mail: srjabchikov@hotmail.com

### Abstract

The role of the monument Moai a Mata Mea at Easter Island in the fertility cult has been explained. The own late statue that belonged to the ruling Miru group can provisionally be dated back to ca. 1691 A.D. on the backbone of the archaeoastronomical studies as well as the written sources. Some data about watchings of the total solar eclipse of September 16, 1773 A.D. and Halley's Comet of 1682 A.D. are presented as well. A new information on the early settlement of Rapanui is of our interest, too.

### Introduction

Easter Island is a remote spot lost in the Pacific Ocean where many ceremonial platforms were oriented toward certain positions of the sun (Mulloy 1961, 1973, 1975; Liller 1991). Therefore numerous platforms and statues could be used as specific astronomical instruments.

### Mars in the Rapanui Beliefs

The monument Moai a Mata Mea stands close to the north of the royal residence Tahai. Its height is around 2.5 m. The statue's name literally means 'The statue (by the name of) *Mata Mea* [The Red Eye],' and this personal name denotes Mars, besides, the grammatical article *a* for proper names introduces it. Why did the natives choose this place for the monolithic sculpture? It is evident that this celestial body played the great role in the religious beliefs of the local rulers.

On a staff belonged initially to the prominent king *Nga Ara* (Rjabchikov 2012: 566-567, figure 3; 2014a: 8, appendix 1) a passage (figure 1) says:

1(I 12): 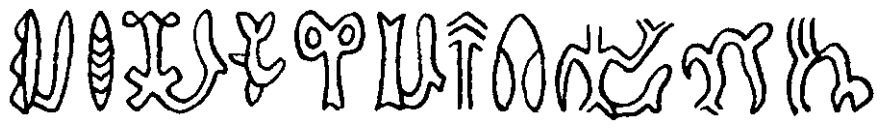

Figure 1.

1 (I 12): **18-4 33 49 (102 123) 60 4 (102) 4-33 64 44var (102) 44b 58 62** *Te atua (h)ua ma(h)ua Mata atua-atua Mea: taha tua tahi to*. 'The lord (king) makes fertile the sides (fields) where the first *tua-tea* sweet potatoes are added (= are planted; grow) because of the great god Mars (*Mata Mea*).'

This text correlates with the beginning of a Rapanui chant about the power of the king (Métraux 1937: 52-54; with my corrections in the left part):

| | |
|---|---|
| *E aha to, rau ariki ki te mahua i uta nei?* | What does the king make fertile in the country? |
| *He tupu, tomo a Matamea i rangi, rau, he tua-tea to, rau ariki ki te mahua i uta nei.* | Mars comes up, appears in the sky. The king makes the shoots of the white sweet potatoes grow in the country. |
| *Anira to, mani roto ka rata te tua-tea,* | Now he makes the sweet potatoes favourable, |
| *ka rata te rangirangi, ka rata te tupuna.* | the sky favourable, the ancestors favourable. |

We know that in the southern hemisphere Mars as well as the sun, the moon and other planets moved and move in the sky from east, trough north, to west. The statue Moai A Mata Mea was a symbol (embodiment) of Mars seen in its upper celestial position from the royal residence Tahai. In my opinion, this Rapanui name of Mars has the Proto-Samoan origin, compare Samoan *Mata-memea* 'Mars,' literally not only 'Red-face' (Stair 1898: 48), but also 'Red-eye.' Mars was called *Matawhero* [*Mata whero*] in New Zealand (Best 1922: 33); this name also signifies 'The Red Eye.' Tahitian name of Mars was *Feti-aura* [*Fetia* = *Fetu* (*'*)*Ura*] 'The Red Star' (Davies 1851: 86). Hawaiian name of Mars was *Hōkū-'ula* [*Hōkū 'ula*] 'The Red Star' (Pukui and Elbert 1986: 76). In the *rongorongo* records on the Keiti (Ev 6) and Small Santiago (Gr 1, Gv 7) tablets the name of this planet is put down as glyphs **60-23** *Mata Ura* 'The Red Eye.'

Here and everywhere else, I use the computer program RedShift Multimedia Astronomy (Maris Multimedia, San Rafael, USA) to look at the heavens above Easter Island. Since the natives planted sweet potatoes in particular in September (Barthel 1978: 52), and sooty terns – the birds of bird-man rituals – arrived in September, one can calculate days when Mars was seen high in the morning sky. For example, on September 15, 1691 A.D. Mars rose at 00:28 a.m., and its transit was at 05:42 a.m. On the other hand, the beginning of dawn was at 04:58 a.m., and the sunrise was at 06:15 a.m. So, from environs of the ceremonial platform Tahai Mars was seen in its upper movement in the north direction.

In the Rapanui language the words signifying 'height' and 'growth' are close cognate: cf. Rapanui *roa* 'long; large; extent,' *roroa* 'elongated; to grow tall' and *roaroa* 'long; tall; to grow; height.'

## Several Parallels in *Rongorongo* Inscriptions

In the records on both sides of the Tahua tablet the total solar eclipse occurred in the morning of September 16, 1773 A.D. is described. Here I offer the reading of a section of the inscription (Ab 7) about this event: at first one can disclose the archaic verbal particles *ha* (the perfect tense). The name of the god *Makemake* (the personification of the sun) is introduced by the particle *ko* for personal names. The Old Rapanui *tara* 'to dawn' has the same form outside of the island: cf. Maori *tara* 'rays of the sun, shafts of light, appearing before sunrise.' This text contains Rapanui *tomo* 'to reach; to land; to penetrate; to enter' used in the song about the power of the king, see above. The inscription (figure 2) reads in this fashion:

1(Ab 7):

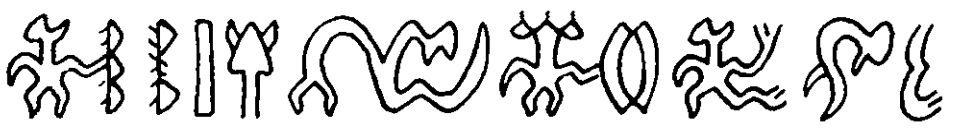

Figure 2.

1 (Ab 7): **6 17-17 4 21 31 5 6 57 6 62-43** *Ha teatea atua ko Make(make)-atua, ha tara, ha tomo*. 'The god *Makemake* (= the sun) appeared, (the sun) peeped out, (it) entered (rose).'

Interestingly, the relationship between the harvest size of sweet potatoes and the religious symbolism (the Frigate Bird; the Seal/Fish as a legendary progenitor; the god-creator *Makemake*) of the ruling tribe Miru is reflected in four clauses as a thanksgiving prayer on the Keiti tablet (Er 7), see figure 3:

1(Er 7):

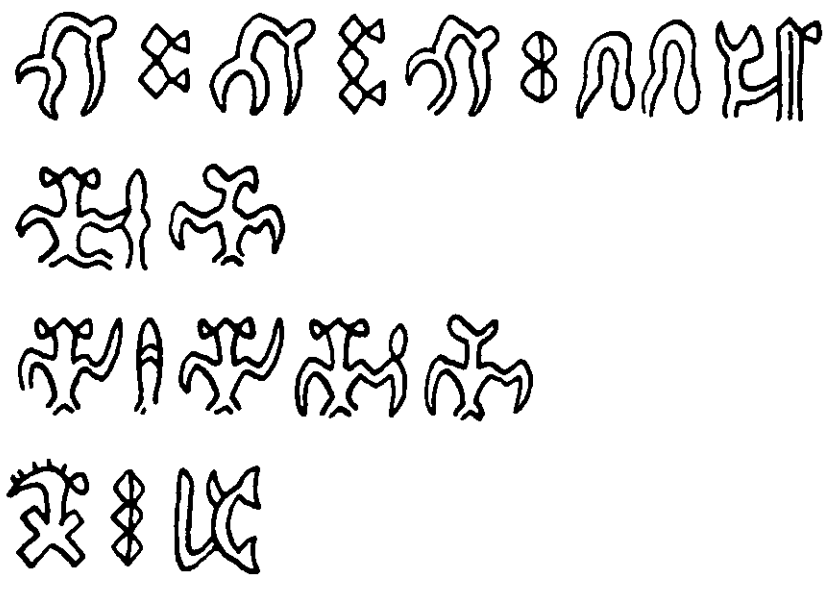

Figure 3.

1 (Er 7): **44b-17 44b-17 44b-17 25-25 19-26** *Tua-tea, tua-tea, tua-tea huahua kumaa*. '(There are) many fruits of the *tua-tea* variety of sweet potatoes (*kumara, kumaa*).'

**69-73 44** *Makohe TAHA*. '(It is a tribal ancestral symbol of the Miru group:) The Frigate Bird.' (Cf. Rapanui *makohe* 'frigate bird'and Mangarevan *mokohe* 'ditto;' it is the phonetic reading of glyph **44**.)
**6 11 (**or **12) 6 30-44** *A Pakia* (or *Ika*), *a Anakena*. 'The Seal or Fish (another symbol of the same group: the god *Tangaroa*), Anakena (one of royal residences; the home of the legendary king *Hotu Matua*).'
**49 17 4-8** *Ma(h)ua te Atua-Matua*. 'The god *Atua-Matua* [the god-Father] (*Tane*, *Tiki* = *Makemake*) makes fertile (all the world, for instance, soil, various fruits, eggs etc.).'

There are grounds to assume that the place name *Tahai* derived from the terms *Taha* 'Frigate Bird' and *(h)i* 'sunbeams' (Rjabchikov 1997a: 208). This bird was an emblem of the Miru (Barthel 1978: 151). The god *Tangaroa* was the head of the Miru kings (Métraux 1940: 127; Lee 1988; Rjabchikov 2014b).

Now one can offer three identical texts (with variations of some signs) inscribed on the Great Santiago (H), Great (P) and Small (Q) St. Petersburg tablets and telling of the fertility which was gained from the patron deities *Haua* and *Makemake*, see figure 4.

1(Hr 7-8):
2(Pr 7):
3(Qr 7):

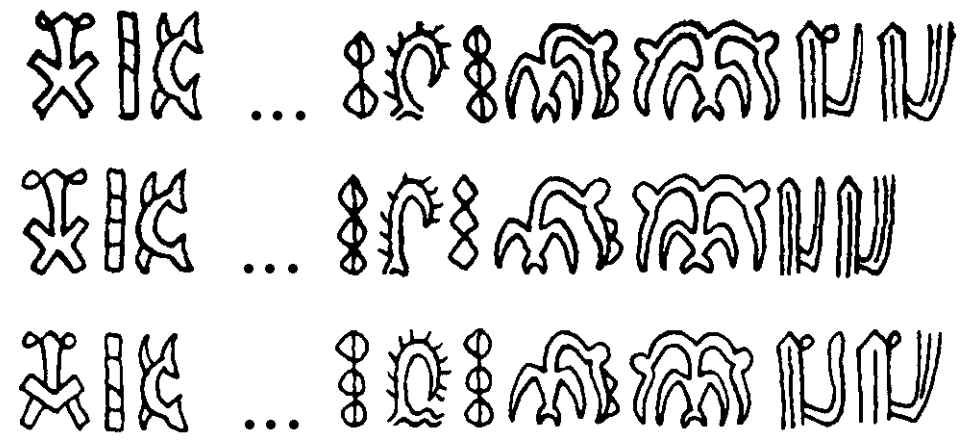

Figure 4.

1 (Hr 7-8): **49 4-8 … 17 14 17 44-44 26-5 26-32** *Ma(h)ua Atua-Matua… Te Haua, te Tahataha-Matua ma(h)ua*. 'The god *Atua-Matua* (*Tane*, *Tiki*, *Makemake*) makes fertile (the world). (The moon goddess) *Haua*, the great Frigate Bird (*Tane*, *Tiki*, *Makemake*, as the sun god)-Father make fertile (the world).'
2 (Pr 7): **49 4-8 … 17 14 17 44-44 26-5 26-32** *Ma(h)ua Atua-Matua… Te Haua, te Tahataha-Matua ma(h)ua*. 'Ditto.'
3 (Qr 7): **49 4-8 … 17 14 17 44-44 26-4 26-32** *Ma(h)ua Atua-Matua… Te Haua, te Tahataha-Matua ma(h)ua*. 'Ditto.'

Here the words *ma(h)ua* and *matua* are taken down as ideograms and combinations of quasi-syllabic signs. The names of the deities are introduced by the grammatical article *te*, and the combinations of this particle and another personal name are rendered twice in the local Creation Chant: they are *te Hina-kauhara* (the moon goddess) and *te Ririkatea* (a royal forefather) (Métraux 1940: 320-322).

### Additional Data about the Red Monument Dedicated to Mars

The statue Moai a Mata Mea was cut not from tuff at the Rano Raraku quarry, but from red scoria at the Puna Pau quarry. The latter name means 'The Cut Well (Source).' These records on the Mamari (C), Small Santiago (G), Small Washington (R) and Great Santiago tablets describe the latter quarry in different religious terms, see figure 5.

1(Cb 2):
2(Gr 2-3):
3(Ra 6):
4(Hv 12):

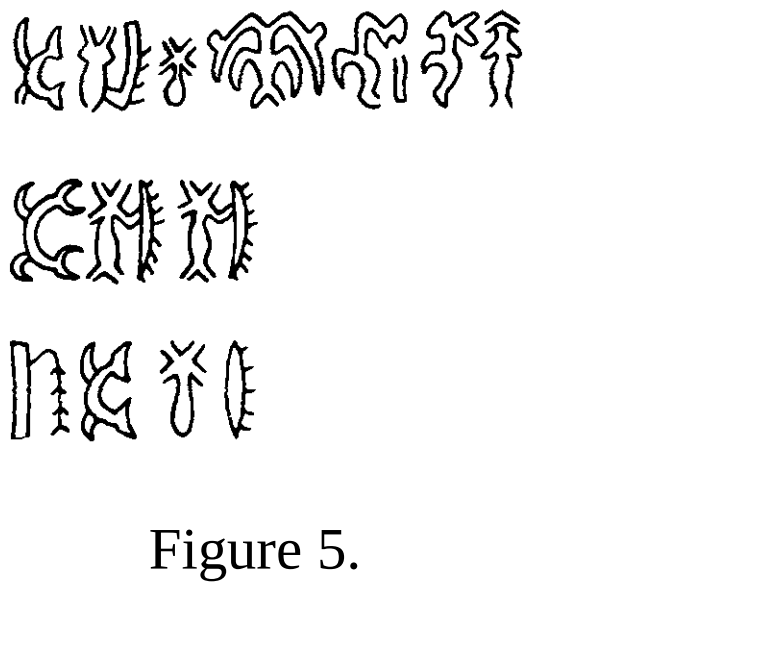

Figure 5.

1 (Cb 12): **6-4 45-46-45 11 19 4 19 59-33 12** *A atua Puna(pu) Pakia ki atua, ki kaua Ika*. 'The god 'The Cut Well [*Puna Pau*] of the Seal [*Tangaroa*] for the deity, for the ancestor (called) Fish [*Tangaroa*]'.'
2 (Gr 2-3): **8 45-46-45 44 62 4 19 12 33** *Motu Puna(pu) Taha to atua ki Ika VAI*. 'The Cut Well [*Puna Pau*] of the Frigate Bird, (there are) additions to gods for the Fish [*Tangaroa*].'
3 (Ra 6): **8 45-46-45-46** *Motu Punapuna*. 'The Cut Well [*Puna Pau*].'
4 (Hv 12): **4 24 8 45-46** *atua ai Motu Puna*. 'The god/location (called) 'The Cut Well' [*Puna Pau*].'

The Mamari tablet was written by a scribe of the Tupa-hotu tribe after the cruel war between the Miru tribe together with the allies (Hanau momoko) and the Tupa-hotu tribe together with the allies (Hanau eepe) (Rjabchikov 1994; 1997b; 2012: 567). The text on this board discussed above in fragment 1 means that the Tupa-hotu men were forced to accept the fact that the Puna Pau quarry belonged to the Miru tribe who worshipped the great god *Tangaroa*.

Let us consider a stone "hat" for a statue manufactured from red scoria of the Puna Pau quarry and lying not far from this site (Lee 1992: 125, figure 4.133). Glyph **12** *Ika* 'The Fish' is depicted among the symbols carved on this "hat," see figure 6, fragment 1. Another sign on that stone cylinder depicting a bird as the head figure of Hotu Matua's legendary ship called *Oraora Ngaru* (Rjabchikov 2011) has successfully been decoded. I presume that the presence of these signs witnesses that the Miru tribe (the adepts of the cult of the sea god *Tangaroa*) possessed the quarry when timber became rare on the island.

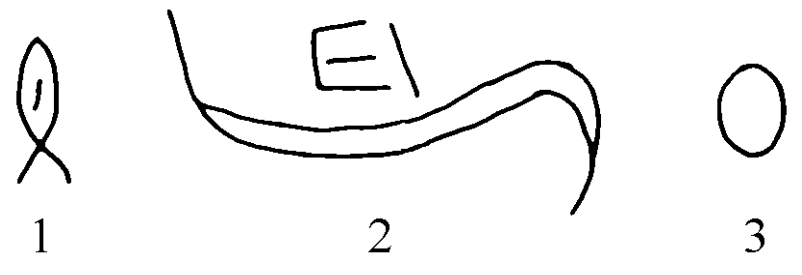

Figure 6.

The most recent date for the ancient works with statues at the Rano Raraku quarry was ca. 1650 A.D. (Skjølsvold 1993: 94). Relying on this, the viewpoint has been proposed that the decadent phase began on the island that year (Esen-Baur 1993: 151). I think that one can get the additional information about that war (the series of wars) on the basis of the folklore texts "Victory of Tuu Foretold," "Matakaroa" and "The Long Ears and the Short Ears" (Métraux 1940: 382-383; Knoche 1912: 873). The god *Hiku-nene-ko-mo-toi-pua* (cf. Rapanui *hiku* 'tail') who was mentioned in the first narration was the designation of Halley's Comet of 1682 A.D.; hence, this event established the time frame of the tremendous war (Rjabchikov 1995: 45). In compliance with this text, the authority of the Miru tribe over the whole island lasted ten years after that victory. Admittedly, the radiocarbon dating (the probability of 2σ) of the K-501 sample (charcoal) from the Poike Ditch gave the wide range from 1450 to 1890 A.D., and for the probability of 1σ was the range from 1490 to 1690 A.D. (Martinsson-Wallin and Crockford 2001: 249, table 1), but indeed such a dating did not deny Englert's (1974) dating of the decisive battle on 1680 A.D. figured out from local genealogies.

Let us consider a large rock picture on a panel beside the ceremonial platform Ahu Raai (Lee 1992: 108-109, figure 4.107). Due to its extraordinary length, the double canoe has a central position here.

On the left above the upper canoe there are such glyphs (figure 6, fragment 2): the big sign **108b** *iri* (to go in a canoe; to elevate) and signs **26-4** *Matua* (= the name of *Hotu Matua*) written in cursive. Thus, the arrival of the first king Hotu Matua of the Miru tribe swimming in a double canoe is described.

Let us count cupules (dots) carved beneath the lower canoe. I have got the 35 dots. According to Barthel (1978: 73, table 4), the voyage of the explorers (the brave young men sent by Hotu Matua) from the legendary homeland Hiva to Rapanui lasted 35 days. So, the 35 dots = the 35 days of the voyage of Hotu Matua with his crew from Mangareva to Easter Island.

The bald head is depicted in the picture to the right of the upper canoe. Glyph **115** *Taka* is inscribed nearby, see figure 6, fragment 3. It is the name of the leader of the triumphant rebels called *Ma-Taka Roa* 'The Great *Ma-Taka* [the Round (head)]' and *Poeu Marengo* [*po e ua Marengo*] 'The Bald-head is holding (*popo*) the royal staff (*ua*)' in the folklore sources.

I believe that the Miru artists created such a masterpiece soon after the great victory of the their tribe over the Tupa-hotu one.

Consider the following record on the Small Washington tablet, see figure 7.

1(Rb 6-7):

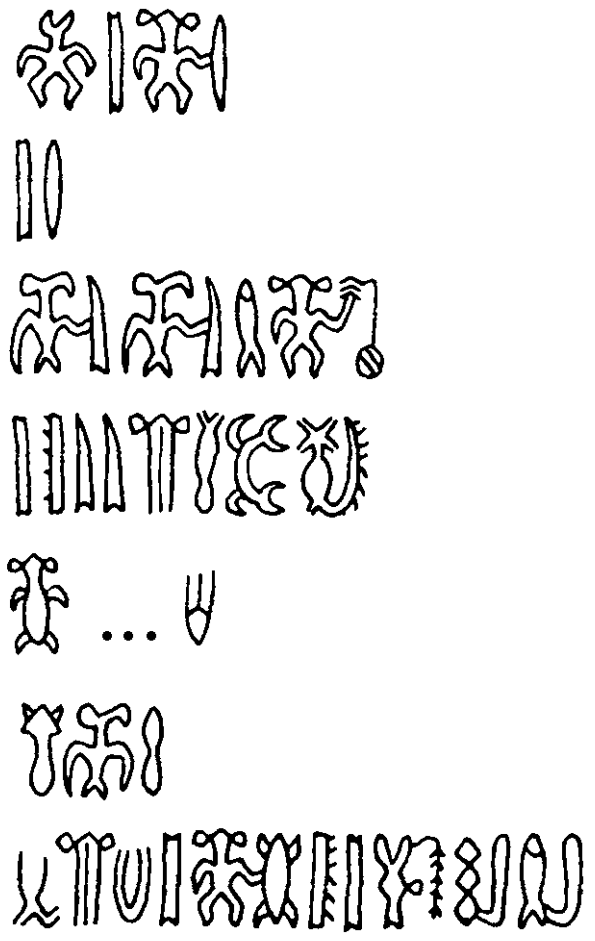

Figure 7.

1 (Rb 6-7): **6 4 6 30** *Ha ati, ha ana.* '(They) had cut numerous (stone hats).'
**4 30** *Ati, ana.* '(They) cut numerous (stone hats).'
**44-5-44-5 12 6-33 103** *Tatutatu Ika hau, pea.* '(They) carved (the sign) 'The Fish' on the (stone) hat.'
**4-4 5-5-26 9 8 45-46** *Atiati tutuma niu Motu Puna.* '(They) felled a coconut palm (to make logs because of the works at) the Puna Pau quarry.' (Perhaps the members of the Miru group hewed alone tall tree down and then attempted transporting that stone cylinder from the quarry.)
**68 7**? **49**?[1] (a damaged segment, 3 or 4 signs) **1** *Hono Tuu* (*ariki*) *mau* … *Tiki.* 'The assembled Tuu people (Miru etc.) of the king (killed the king of the god) *Tiki* (the emblem of the Tupa-hotu tribe).'
**21 44 65** *Ko Ta(h)a rangi.* 'The Frigate Bird (the emblem of the Miru tribe) sent (the rebels).' (Cf. Samoan *'ātafa* 'frigate bird' and Tahitian *otaha* 'ditto' < **kotaha* 'ditto.')
**48 26 115 4 6-28 4-4 54 24 17-4 12 4** *Ua Ma-Taka-atua Hanga atuaatua kai-ai te atua Ika-atua…* '(It was) the authority (the royal staff *ua* literally) of the lord *Ma-Taka* from the Bay of the great god of the land (*kainga*, *ai*) of the god 'The Fish' (= *Tangaroa*) (= from the Anakena bay)…' (The apexes of glyphs **48** *u, ua,* **26** *maa, ma, mo* and **115** *taka* are badly eroded.)

According to Routledge (1998: 199), red stone cylinders for statues were called *hau* (hats) *hitirau moai* (a hat of the stone called *hitirau* due to (for) a statue literally).

### Again about the Tahai Ceremonial Complex

This territory includes the ceremonial platform Ahu Tahai with a single statue, the ceremonial platform Ahu Vai Uri [The ceremonial platform of the Black Water] with five statues (earlier with six ones) and the ceremonial platform Ahu ko te Riku with a single statue.

All the three platforms are described in religious terms in a record (figure 8) on the Keiti tablet.

1(Ev 5):

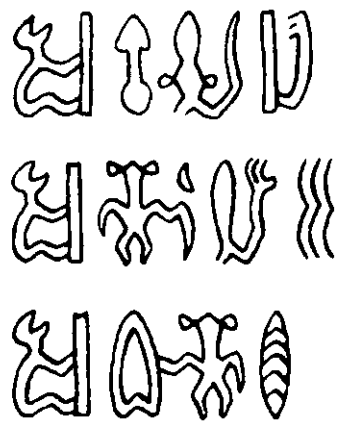

Figure 8.

[1] See a photograph: <http://collections.si.edu/search/tag/tagDoc.htm?recordID=nmnhanthropology_8010183>.

1 (Ev 5): **6-4 22 73 50 4-15** *A atua ao (h)e (H)i; atua roa;* ‘The deity ‘The authority of the Sunbeams’ (the great god);’

**6-4 6 3 48-15 33** *a atua a Hina Uri Vai;* ‘the deity ‘The Moon of the Blackness of the Water;’’

**6-4 47 6-33** *a atua Ava Haua*. ‘the deity ‘*Ava* (*Avae*, the Full Moon)-*Haua*.’’

Reasoning from this text, the platform Ahu Vai Uri was a lunar observatory. All the statues were the personifications of different phases of the moon.

The clue to the name of the platform Ahu ko te Riku is Rapanui *riku* ‘to grow in abundance.’ According to the read text, the last platform was a part of that observatory and its statue was an image of the (almost) full moon. Remember the story about the legendary hero *Tuu-ko-Iho* (Métraux 1940: 260-261). One day he went from the platform Ahu-te-peu to the bay Hanga Hahave past the platform Ahu-ava-a-atea. The last site has not been identified yet. *Tuu-ko-Iho* could walk through the Tahai area. Therefore I suppose that the name of *Ava-a-atea* [The white full moon] = the name of *Ava Haua* from the inscription = the modern name of *ko te Riku* [The abundant or full (moon)].

Moreover, the platform Tahai and its statue (the image of the sun) could serve as a solar observatory (the constant watchings of the setting sun and other celestial bodies in the direction of the western ancestral homeland called *Hiva* or *Havaiki*). The wordplay in the terminology is also conceivable: cf. Rapanui *taha* ‘to set (of the sun).’ The two other platforms and their statues were certain components of the big observatory.

## Conclusions

The obtained results suggest that the statue Moai a Mata Mea was cut and erected in September 1691 A.D. (when Mars was seen high in north from the royal residence Tahai) or shortly thereafter. A stone head made of red scoria, which was discovered in the sea at Tahai behind Ahu Vai Uri, seems to date from the same time. This artefact was the image of Mars gone beyond the horizon.